\documentclass[journal]{IEEEtran}
\usepackage[export]{adjustbox}  
\usepackage{color,soul} 

\usepackage{jabbrv}
\usepackage{cite}

\ifCLASSINFOpdf
\else
\fi

\begin{document}
%
\title{Stepped-Frequency THz-wave Signal Generation From a Kerr Microresonator Soliton Comb }
%
%
%

\author{Omnia Nawwar,
        Kaoru Minoshima,~\IEEEmembership{Member,~IEEE, Fellow,~Optica,}
        and~Naoya Kuse,~\IEEEmembership{Member,~Optica}

\thanks{This work was supported by Precursory Research for Embryonic Science and Technology
(JPMJPR1905); Japan Society for the Promotion of Science (21H01848,
21K18726); Cabinet Office, Government of Japan (Subsidy for Reg. Univ.
and Reg. Ind. Creation).}
\thanks{O. Nawwar, K. Minoshima, and N. Kuse are with Institute of Post-LED Photonics, Tokushima University, Tokushima, Tokushima, 770-8506, Japan.  K. Minoshima are also with Graduate School of Informatics and Engineering, The University of Electro-Communications, Chofu, Tokyo, 182-8585, Japan. (Corresponding author: Naoya Kuse. email: kuse.naoya@tokushima-u.ac.jp).}
\thanks{Manuscript received XXX; revised XXX.}}

%
%

\markboth{Journal of \LaTeX\ Class Files,~Vol.~, No.~, month~}%
{Shell \MakeLowercase{\textit{et al.}}: Bare Demo of IEEEtran.cls for IEEE Journals}
%



\maketitle

\begin{abstract}
Optically generated terahertz (THz) oscillators have garnered considerable attention in recent years due to their potential for wide tunability and low phase noise. Here, for the first time, a dissipative Kerr microresonator soliton comb (DKS), which is inherently in a low noise state, is utilized to produce a stepped-frequency THz signal ($\approx$ 280 GHz). The frequency of one comb mode from a DKS is scanned through an optical-recirculating frequency-shifting loop (ORFSL) which induces a predetermined frequency step onto the carrier frequency. The scanned signal is subsequently heterodyned with an adjacent comb mode, generating a THz signal in a frequency range that is determined by the repetition frequency of the DKS. The proposed method is proved by proof-of-concept experiments with MHz level electronics, showing a bandwidth of 4.15 GHz with a frequency step of 83 MHz and a period of 16 $\mu$s.  
\end{abstract}


%
\IEEEpeerreviewmaketitle

\section{Introduction}
%
%
%
%

\begin{figure*}[!ht]
\centering
\fbox{\includegraphics[width=0.95\linewidth]{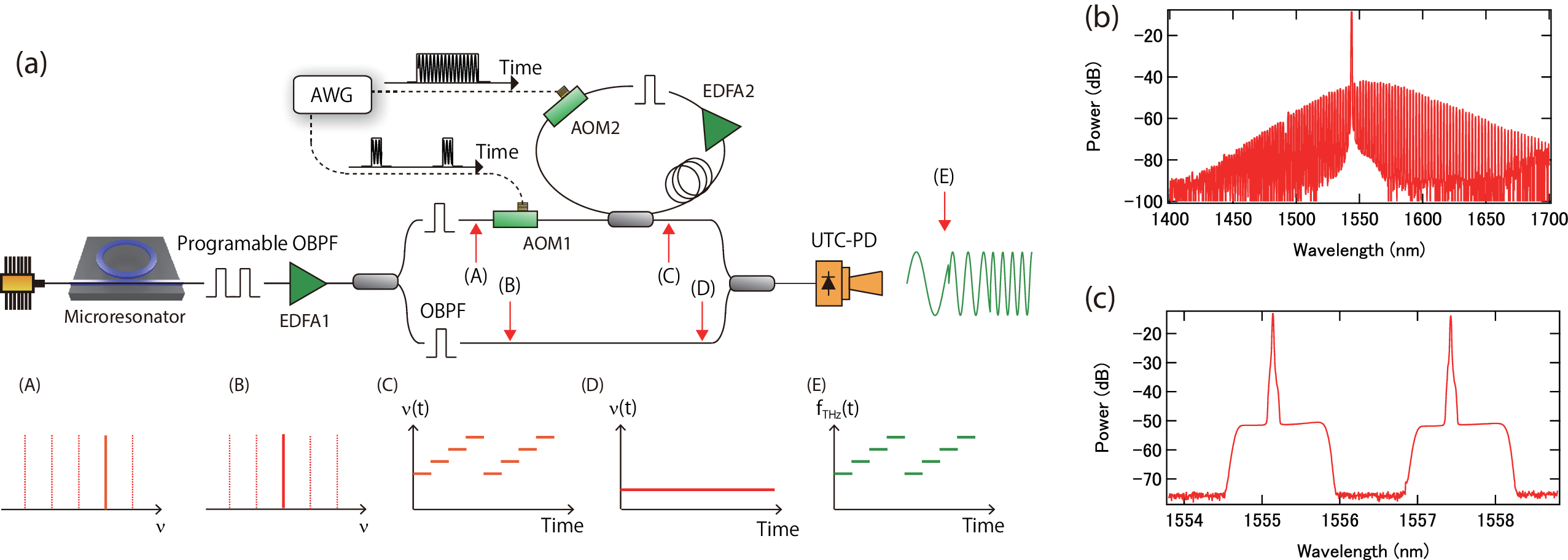}}
\caption{(a) Schematic of the experimental setup. AWG: arbitary waveform generator, EDFA: Er-doped fiber amplifier, AOM: acousto-optic modulator, OBPF: optical bandpass filter, UTC-PD: uni-travelling-carrier photodiode. (A) and (B) show illustrations of the comb modes at locations (A) and (B). (C), (D) and (E) show the instantaneous frequency of the comb mode/THz at locations (C), (D) and (E). (b) Optical spectrum of the DKS. (c) Optical spectrum of the filtered comb modes after the programmable OBPF.}
\end{figure*}

\IEEEPARstart{T}{he} potential applications of terahertz (THz) signals (0.1-10 THz) have experienced substantial expansion in recent years, fostering an escalating interest in the advancement of THz technologies \cite{Banks_2023,Chen_2022,Nagatsuma_2016}. Specifically, the necessity for wideband or frequency-scanned THz is unequivocal in applications to radar \cite{Cooper_2011}. Generation of THz signals can be principally accomplished via two techniques - electronically or photonically. Electronic methods, wherein microwave frequencies are multiplied to the THz domain, introduce inherent obstacles due to nonlinearity, parasitic effects, and a decline in both system efficiency and noise level, further complicated by fabrication difficulties associated with achieving nanoscale precision. Alternatively, photonic generation of THz signals has emerged as an advantageous substitute, where optical frequency is downconverted into the THz domain. Photonic approaches can be subdivided into two categories: photomixing of a single frequency continuous-wave (CW) laser with a chirped-mode locked laser \cite{Lin2005, Rashidinejad2014}, and photomixing of two optical tones separated by the desired THz frequency \cite{Danion_2014,Kittlaus2021,hirata2005low,Jerez_2019,parriaux2020electro,Jia_2022,Zhang_2019,tetsumoto2021optically,Kuse_2022}. As the former is encumbered by complexity and limited reconfigurability (e.g., chirp rate), the latter has been extensively adopted. For THz wave scanning, the frequency of one of the two optical tones is varied. A simple strategy is to use two single-frequency CW lasers and directly modulate the frequency of one \cite{Danion_2014,Kittlaus2021}, but this transfers the CW lasers' phase noise to the generated THz wave, thereby degrading the sensitivity of radar systems and hindering to reach scan-bandwidth-limited resolution due to non-ideal linearity and reproducibility \cite{ahn2005suppression}. An alternative approach to THz signal generation is through the use of two optical tones extracted from electro-optic frequency combs \cite{hirata2005low,Jerez_2019,parriaux2020electro,Jia_2022}. In this context, the stability of the produced THz wave is only constrained by the reference microwave signal, surpassing the performance of two independent CW lasers. Nonetheless, the necessity for high-bandwidth electro-optic modulators (EOMs), high-power microwave amplifiers, and a frequency-tunable microwave oscillator to actuate the EOMs, convolute the entire system. In addition, the scan range is restricted by the comb mode spacing (typically, 10 – 20 GHz) due to optical bandpass filters (OBPFs) employed to isolate the two comb modes.

Recently, dissipative Kerr microresonator soliton combs (DKSs) \cite{Herr_soliton,kippenberg2018dissipative} have been the focus of considerable attention as a method for THz wave generation \cite{Zhang_2019,tetsumoto2021optically,Kuse_2022}. DKSs are generated by inputting a single-frequency CW laser into high-Q microresonator, which are fabricated by CMOS-compatible processes \cite{kovach2020emerging}, thereby rendering them a chip-scale, mass-producible laser source \cite{xiang2021laser}. Additionally, the comb mode spacing of DKSs ranges from 10 GHz – 1 THz, which is suitable for the generation of THz waves. Furthermore, DKSs are in a mode-locked state, showing high coherence among the comb modes. Along with locking the repetition frequency of DKSs to external references such as a Brillouin cavity \cite{tetsumoto2021optically} and fiber delay lines \cite{kwon2022ultrastable, Kuse_2022}, THz waves generated from DKSs exhibits ultra-low phase noise with -100 dBc/Hz at a 10 kHz frequency offset for 300 and 560 GHz carriers \cite{tetsumoto2021optically,Kuse_2022}. However, the repetition frequency of DKSs comb cannot be largely scanned. Even with a microheater deposited on a micresonator, the scan range is 0.1 \% of the repetition frequency \cite{Kuse_2020,kuse2021frequency}, which corresponds to the frequency scanning of a 300 GHz wave of as small as 30 MHz, prohibiting the use of THz wave generated from DKSs for radar. 

In this study, we propose and demonstrate a method to scan the frequency of a THz wave generated from a DKS. Our proposed method extracts two neighboring comb modes from a DKS and scans the frequency of one of these comb modes using an optical recirculating frequency-shifting loop (ORFSL) \cite{Liu_2022,Chen_2018,Lyu_2022,Zhang_2020multioctave,Schn_belin_2019,Zhang_2022}. Upon heterodyning the two comb modes at a uni-traveling-carrier photodiode (UTC-PD), a frequency-scanned THz wave is produced \cite{Ito_2005}. Our proof-of-concept experiment effectively scans the frequency of a THz wave generated from a DKS from 278.7 to 282. 8 GHz in 16 $\mu$s. This corresponds to a bandwidth of 4.15 GHz with a frequency step of 83 MHz. 

\section{Experimental setup and operation principle}
A basic architecture of the experimental setup and signal at different points are depicted in Fig. 1(a). The output of a pump CW laser is modulated by a dual-parallel Mach-Zehnder modulator (DP-MZM) (not shown in Fig. 1(a)), amplified by an Er-doped fiber amplifier (EDFA) (not shown in Fig. 1(a)), and coupled into a high-Q Si$_3$N$_4$ microresonator (Ligentec SA) with a free-spectral range of about 280 GHz. The DP-MZM is operated in a carrier-suppressed single-sideband (CS-SSB) mode, which is used to rapidly scan the frequency of the pump CW laser to access a stable DKS \cite{stone2018thermal}. More details to generate a DKS is described in the appendix and ref \cite{kuse2019control}. The optical spectrum of the DKS used in this work is shown in Fig. 1(b). The comb mode spacing and 10-dB bandwidth are about 280 GHz and 90 nm, respectively. 

A bandstop filter (not shown in Fig. 1(a)) is used to reject the residual pump CW laser before feeding the DKS to a programmable OBPF to pass a pair of neighboring comb modes used in our experiment. The two neighboring comb modes at the wavelengths of 1557.44 and 1555.14 nm, corresponding to -6th and -5th with respect to the pump mode, are shown in Fig. 1(c). The signal-to-noise ratio (SNR) of the comb modes is about 38 dB with the resolution bandwidth (RBW) of 0.02 nm. Extracted modes are amplified to about 60 mW before splitting into two branches through a 50/50 optical coupler. Individual comb modes are extracted by OBPFs in the upper and lower branches as shown in Fig. 1(a) at locations (A) and (B). The signal in the upper branch experiences the frequency shifts by an ORFSL \cite{shimizu1992technique}. The ORFSL consists of an acousto-optic modulator (AOM) (AOM 1 in Fig. 1(a)) with an extinction ratio of more than 50 dB and a frequency shifting loop (FSL). AOM 1, which is drived by a RF signal from an arbitrary waveform generator (AWG), works as an optical switch to convert the comb mode into optical pulses. The pulse width equals to the time delay provided by the ORFSL, and the pulse cycle is determined by the number of round trips (N) of the ORFSL. 

The generated pulses are directed into the ORFSL which is realized by a 50/50 optical coupler, fiber delay, EDFA 2, OBPF, AOM 2, and polarization controller. 
Apart from a 8 cm fiber on the polarization controller, fibers and fiber components in the ORFSL are polarization maintaining. The optical coupler is used as input and output ports of the ORFSL. The length of the fiber used is 49 m which determines the time delay in the loop. EDFA 2 compensates the loss in the loop (mainly due to coupling and insertion losses for components) while the following BPF suppresses the amplified spontaneous emission (ASE) noise. 
AOM 2 shifts the pulse frequency for every round trip by a frequency step of $\Delta f$ (= 83 MHz in our experiment), which corresponds to the driving frequency from the AWG, ensuring the time-frequency linearity of resultant stepped-frequency THz signal. 
When AOM 2 is activated, the input pulse successively experiences the frequency shift of $\Delta f$ every round trips. AOM 2 is deactivated when a new incoming pulse from AOM 1 enters the ORFSL, thereby initiating a fresh cycle of stepped-frequency signal generation process. Note that the driving signals of AOMs 1 and 2 are synchronized so that the timing of the operation of AOMs 1 and 2 are precisely controlled. For each cycle, the optical pulse experiences a frequency shift of $N \times \Delta f$ as shown in Fig. 1(a) at location (C).  Another 50/50 optical coupler is used to combine the scanned signal from the upper branch and the comb mode from the lower branch, whose instantaneous frequency is constant at location (D). Then, it is further amplified by an EDFA (not shown in Fig. 1(a) to provide about 20 mw optical power to UTC-PD  (IOD-PMJ-13001, 280 - 380 GHz). 

Owing to the square-law envelope detection at the UTC-PD, the UTC-PD generates a frequency-stepped THz signal by down-converting the optical frequencies of the comb modes to a THz wave as shown in Fig. 1(a) at location (E). The frequency of the THz wave corresponds to the spacing between the scanned comb mode and the neighboring comb mode, resulting in the scan range of $N \times \Delta f$. For the following experimental results, the time delay and frequency step are kept the same.

\section{Experimental results}
First, we examine a single tone THz signal generated from either two comb modes or two free-running CW lasers without implementing the ORFSL. Our investigation primarily focuses on the SNR and power of the THz signals. The produced THz signal is characterized by down-converting the THz signal to a microwave using the setup shown in Fig. 2(a). The THz from the UTC-PD is mixed with a  frequency-multiplied ($\times$ 12) local oscillator (11.79 GHz) followed by an amplifier (the frequency multiplier and amplifier are incorporated into a single device, WR6.5AMC-I from Virginia Diodes, Inc.) at a sub-harmonic mixer (SHM, WR3.4SHM from Virginia Diodes, Inc.). The down-converted signal is then further amplified by microwave amplifiers and digitally detected by an electrical spectrum analyzer (ESA). Figure 2 (b) shows the relationship between the optical power to the UTC-PD and RF power after down conversion. The THz power (proportional to RF power) is shown to be proportional to the square of the optical input power, mirroring the principle of standard square-law envelope PDs. Although we don’t measure the THz power directly, it is estimated to be 10 $\mu$W according to the datasheet, which could potentially be increased up to 100 $\mu$W by augmenting the optical input power. There is a little difference in power between the down-converted RF signals from the two comb modes and two free-running CW lasers. There is a minor difference in power between the down-converted RF signals originating from the two comb modes and the two free-running CW lasers. This discrepancy is not fundamental, but likely stems from measurement errors or system reproducibility issues such as polarization and temperature dependence of the UTC-PD response. Indeed, the photocurrent of the UTC-PD shows a slight variation from day to day, even with consistent optical input power. The RF spectra of the signals from two CW lasers and comb modes are depicted in Figs. 2(c) and (d). Both cases exhibit almost the same power and noise floor, suggesting that despite the optical SNR of the comb modes being lower than that of the two CW lasers, the quality (i.e., power and SNR) of the photonically generated THz signal is not compromised by using the comb modes. Currently, the noise floor of -65 dBm is limited by the leakage of the local oscillator from the SHM.

\begin{figure}[!ht]
\centering
\fbox{\includegraphics[width=0.95\linewidth]{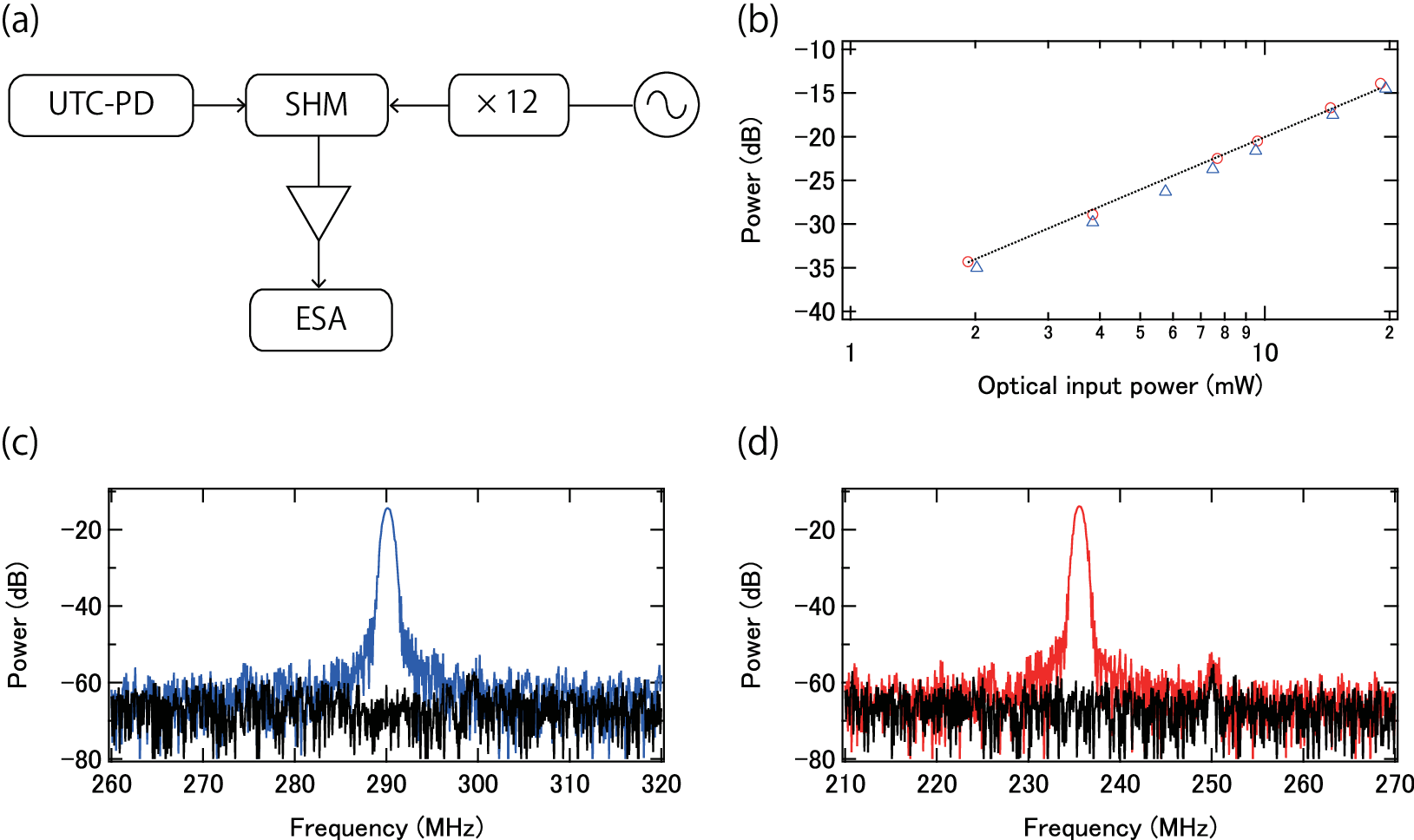}}
\caption{(a) Schematic of the system to down convert a THz signal to a microwave. SHM: sub-harmonic mixer, ESA: electrical spectrum analyzer. (b) Power of down-converted signal when two CW lasers (blue triangle) and two comb modes (red circle) are used. The dotted line show a fitting with a square function to the result of the power when two comb modes are used. (c) RF spectrum of the down-converted signal (blue curve) and noise floor (black curve) when two CW lasers are used. (d) RF spectrum of the down-converted signal (blue curve) and noise floor (black curve) when two comb modes are used.}
\end{figure}

\begin{figure}[!b]
\centering
\fbox{\includegraphics[width=0.95\linewidth]{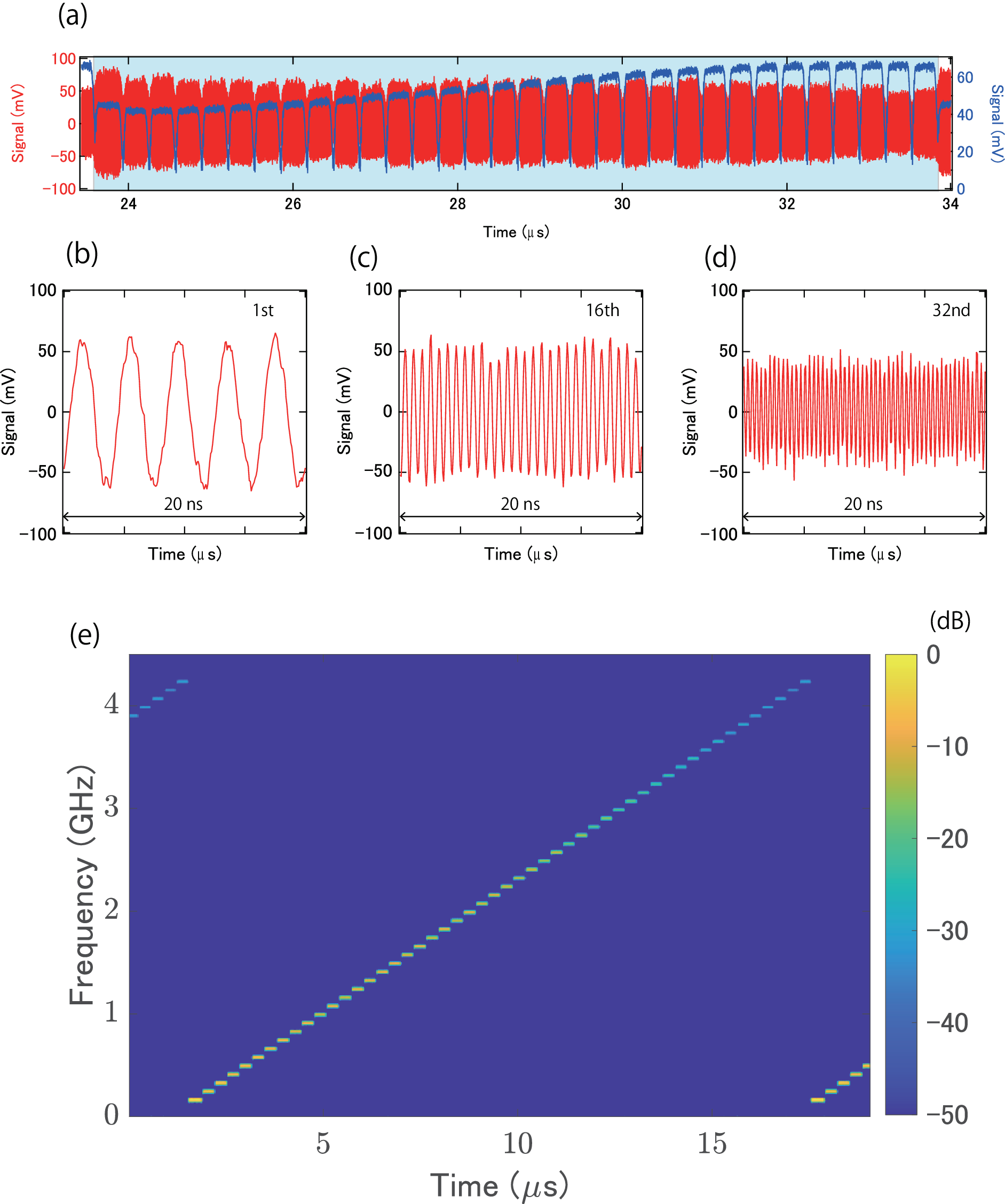}}
\caption{(a) (Blue curve) The intensity of the outputs from the ORFSL. (Red curve) The down-converted signal from a generated THz when the number of round trips in the ORFSL is 32. One cycle is highlighted by the blue square. (b)(c)(d) FFT of the down-converted signal at 1st, 16th, and 32nd, respectively, showing the instanteneous frequency of 0.165 GHz, 1.41 GHz, and 2.738 GHz, respectively. (e) Spectrogram of the down-converted signal when the number of round trips in the ORFSL is 50.}
\end{figure}

Next, we turn our attention to the generation of step-frequency signals utilizing two comb modes. The comb modes employed in this process are extracted by the programmable OBPF as depicted in Fig. 1(c). The mode with a higher frequency is channeled into the ORFSL with the period of the RF signal to the AOMs set to 10.24 µs (equivalent to 32 round trips) and a pulse width of 320 ns. This pulse width is intended to match the time delay within the ORFSL and is experimentally adjusted to minimize the intensity disparity across time slots. 
The intensity of the output from the ORFSL is measured by subtly coupling out the light after the ORFSL (not shown in Fig. 1(a)), which is directed to a PD (not shown in Fig. 1(a)). Blue curve in Fig. 3(a) represents the optical pulses detected after the ORFSL with the PD while the red curve depicts the heterodyned signal detected with the UTC-PD and subsequently down-converted to be displayed on a fast oscilloscope. Given the frequency's gradual increase from the initial to the final loop, later pulses undergo greater attenuation than their leading counterparts due to the frequency-dependent responses of THz components. To counterbalance this attenuation, the intensity of the optical pulses preceding the UTC-PD is escalated, as measured by the PD and displayed as the blue curve in Fig. 3(a). This adjustment is achieved by modulating the pump current of EDFA 2 and the polarization controller within the ORFSL. Figures 3(b), (c), and (d) illustrate the 1st, 16th, and 32nd frequency-shifting pulses within the time domain, displaying a clear frequency increase. The frequency of these bursts (post-down conversion) is calculated using a Fast Fourier Transform (FFT) to be 0.165 GHz, 1.41 GHz, and 2.738 GHz, respectively, each with a frequency step of 83 MHz, thereby yielding a signal with a bandwidth of 2.537 GHz.
Figure 4 presents a spectrogram showcasing the instantaneous frequency of all pulses when a total of 50 round trips (corresponding to a period of 16 $\mu$s) are utilized. The dwell time within one time slot remains at 320 ns, as dictated by the time delay within the loop. The instantaneous frequency of the steps consistently increases with a frequency step of 83 MHz from 0.166 MHz to 4.233 GHz. When considering the frequency down-conversion, the THz frequency is scanned from 278.727 GHz to 282.793 GHz, achieving a bandwidth of 4.1 GHz. 

Lastly, we explore the feasibility of accommodating additional round trips. Two primary concerns arise. First, there's the potential escalation of the system's noise floor, which could be induced by the accumulated noise from the EDFA, coupled with the minimal suppression of the unwanted carrier from the AOM inside the loop. Second, there's the risk of amplitude fluctuations, likely due to the accumulation of imperfect polarization maintaining procedures, paired with the presence of polarization-dependent components in the loop, such as an isolator and Er fiber. Figure 4 shows FFT signals of a single time slot at the 1st, 16th, and 32nd rounds. It is noteworthy that the noise floor does not escalate with an increased number of round trips. In this context, the RF power is not a concern as it is solely determined by the optical input power to the UTC-PD. This outcome could be specific to the THz system, where the noise floor is constrained by the detection setup rather than by the optical tones. Indeed, the degradation of the SNR has been observed in the system for the stepped-frequency microwave using an ORFSL, as mentioned in reference \cite{Liu_2022,Lyu_2022}. 

\begin{figure}[!ht]
\centering
\fbox{\includegraphics[width=0.95\linewidth]{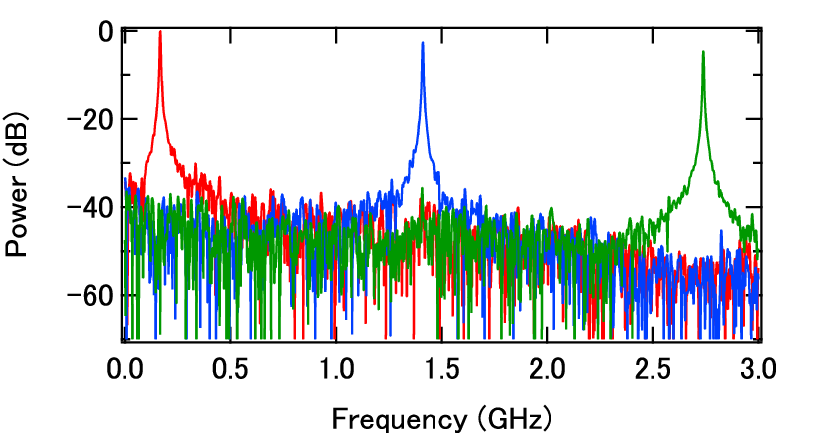}}
\caption{RF spectra of FFT signals of a down-converted signal at 1st, 16th, and 32nd time slots when the number of round trips is 32.}
\end{figure}

While it would be beneficial to observe the noise floor across a larger number of round trips, the limited bandwidth of the equipment (such as the oscilloscope and microwave amplifier) available in our lab precludes conducting the same experiments with more round trips. However, we can assess amplitude fluctuation without generating THz waves by measuring the intensity of the output from the ORFSL. Optical pulses for 100 and 400 round trips are illustrated in Figs. 5(a) and (b), respectively. With a total of 100 round trips, the intensity can be equalized by carefully adjusting the pump current of the EDFA and the polarization controller within the ORFSL. However, managing the optical output power from the loop becomes exceedingly challenging when the total number of round trips reaches 400. The scan bandwidth of the THz is anticipated to be 8.3 GHz for 100 round trips, and 33.2 GHz for 400 round trips, respectively.

\begin{figure}[!h]
\centering
\fbox{\includegraphics[width=0.8\linewidth]{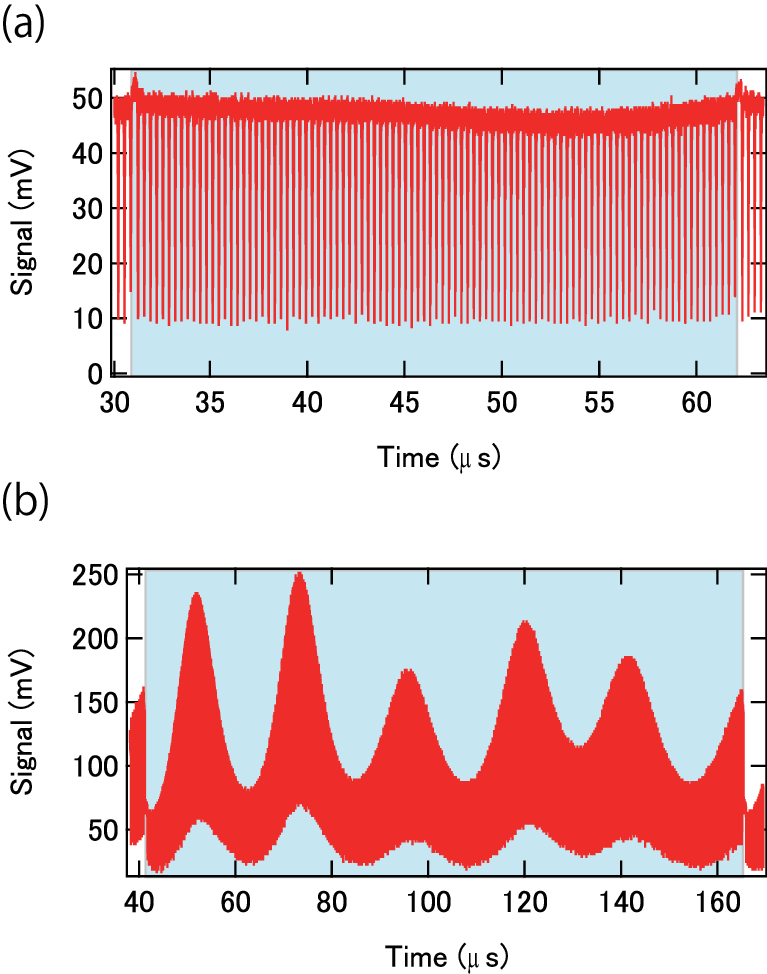}}
\caption{(a) and (b) Intensity of the outputs from the ORFSL when the number of round trips are 100 and 400, respectively. One cycle is highlighted by the blue square.}
\end{figure}

\section{Discussions and conclusion}
While we maintain a fixed time delay and frequency step in the experiments, both parameters can be adjusted. The frequency step of the ORFSL, for instance, can be altered from several MHz to tens of GHz. This can be achieved by choosing an AOM with a different modulation frequency, cascading multiple AOMs, or using a DP-MZM operating in the carrier-suppressed single-sideband mode \cite{Zhang_2020multioctave}. As a result, the ORFSL offers a versatile frequency step capability, adaptable to meet the specific spectral resolution requirements of various applications. Similarly, the dwell time of the step-frequency THz signal within a single time slot can be adjusted by altering the length of the fiber loop, which consequently controls the pulse width. A pulse width of a few nanoseconds can be introduced to achieve shorter dwell times. Notably, since the frequency step and dwell time are governed by separate experimental components, these parameters can be independently optimized, offering a high degree of design flexibility. Furthermore, the number of round trips can be controlled by simply adjusting the driving RF signals to the AOMs, resulting in changes to the bandwidth of the step-frequency THz signals. However, such escalations are not unlimited. Theoretically, while power loss incurred during round trips can be offset by intra-loop EDFA, the SNR continuously decreases until it reaches an unacceptable level. Nevertheless, the SNR of the two modes from the DKSs does not limit the quality of the generated THz, at least up to the 32nd round trip. The SNR of the comb modes can be improved by utilizing the injection locking of DFB lasers to the comb modes, without sacrificing phase noise and frequency stability \cite{Kuse_2021}. More round trips also lead to amplitude fluctuations, as observed in Fig. 5(b). One strategy to counteract these fluctuations involves installing feedback loops to stabilize pulse energy from the loop by controlling the modulation amplitude of AOM 2 and the pump current of EDFA 2. Alternatively, the observed amplitude fluctuations can be used to normalize the THz power by simply dividing by the square of the amplitude fluctuation, though in practice, additional minor modification factors may be necessary. 

In conclusion, we successfully generated a step-frequency-THz signal from DKSs, in which one comb mode's frequency was discretely scanned using an ORFSL, and then heterodyned with a neighboring comb mode. Our experiment involved an ORFSL with 32 and 50 frequency steps (one frequency step = 83 MHz), generating a THz signal with a start frequency of 278.7 GHz and a bandwidth of 4.1 GHz. We also investigated the feasibility of increasing the number of ORFSL round trips, with a particular focus on the SNR and amplitude fluctuation. DKSs demonstrate a high degree of phase correlation between the comb modes, resulting in reduced phase noise distortion and frequency fluctuation of THz signals. The developed system holds potential for application in THz radar \cite{Wang_2022,Liu_2022}, which could be pivotal technology in the era of beyond 5G or 6G.

\appendices
\section{Detailed experimental setup}
\subsection{DKS generation}
A pump CW laser operating at wavelength of 1543 nm is utilized. To rapidly scan the frequency of the pump CW laser, the CW laser passes through a DP-MZM. The DP-MZM is modulated by a voltage-controlled oscillator (VCO). The output of the VCO is amplified and then divided into two signals with a 90-degree phase difference using a 90-degree splitter. These split RF signals are subsequently applied to the DP-MZM. Once the DC biases for the DP-MZM are properly adjusted, the DP-MZM operates in the CS-SSB mode. By altering the frequency of the VCO, the DP-MZM functions as a frequency shifter. The frequency of the VCO is changed by a few GHz from 10 GHz in less than 100 ns. The output from the DP-MZM is then amplified using an EDFA, followed by an OBPF to eliminate ASE from the EDFA. Finally, the pump CW laser, with a power level of about 300 mW, is coupled into a chip with the a microresonator, experiencing a coupling loss of \verb|<| 3 dB.

\subsection{Clock signal generation}
The RF signals applied to the AOMs are produced by AWGs and RF switches. The AWG generates two outputs: a sine wave with a frequency suitable for the AOMs (approximately 80 MHz in our case) and a pulse train. The pulse width and period are determined by the time delay and the number of round trips in the FSL. These sine wave and pulse train are fed into an RF switch, which generates a pulse train with a carrier frequency that corresponds to the frequeny of the sine wave.

\begin{figure*}[!ht]
\centering
\fbox{\includegraphics[width=0.95\linewidth]{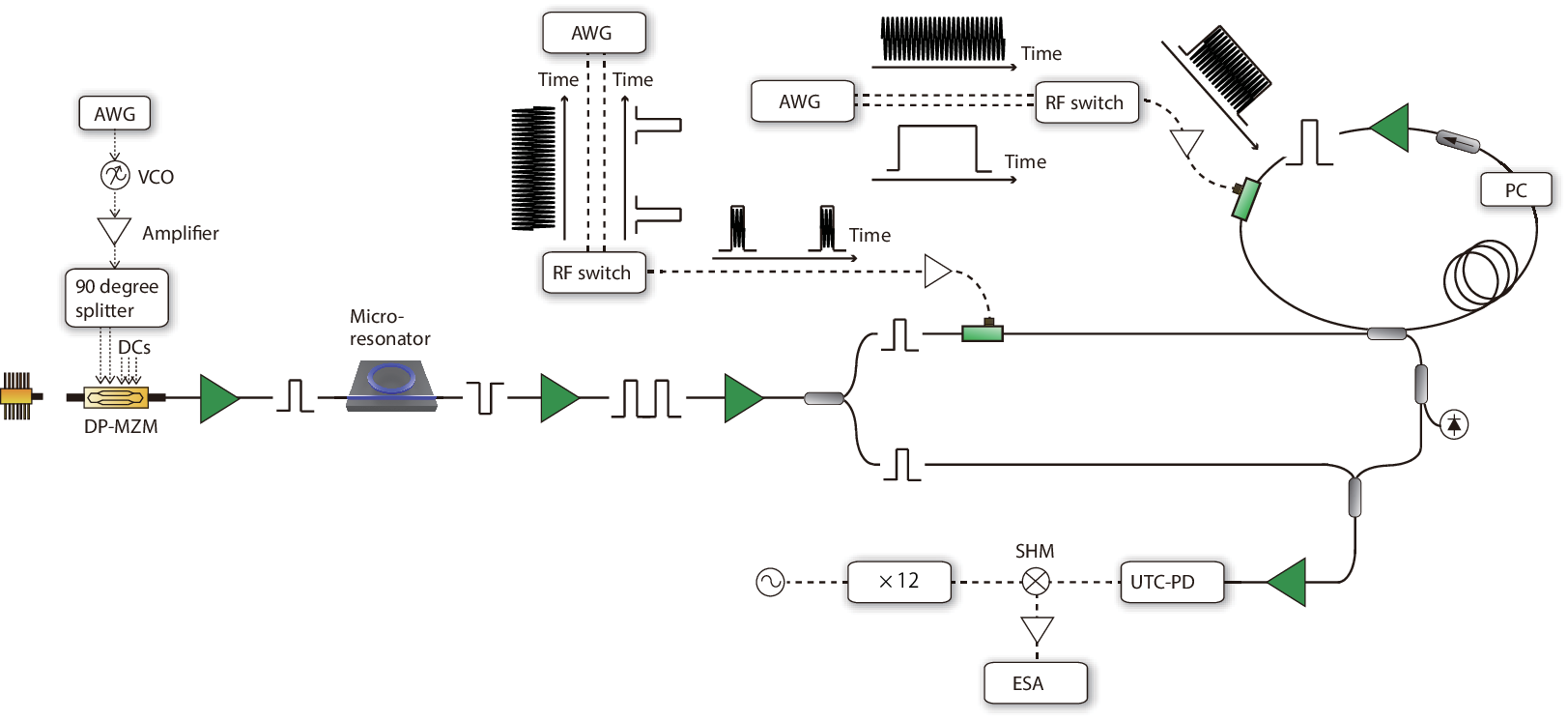}}
\caption{Schematic of the experimental setup. VCO: voltage-controlled oscillator.}
\end{figure*}

\section{Micro/mm/THz wave generation using ORFSL}
The main claim of this study is the generation of frequency-stepped THz signal from a DKS. However, we believe it is instructive to revisit the literature on the generation of micro/mm/THz waves generation using ORFSLs.
Table I presents a summary of the essential parameters. For microwave (\verb|<| 30 GHz) generation \cite{Lyu_2022,Schn_belin_2019,Zhang_2022,Zhang_2020multioctave}, a single cw laser is utilized. The CW laser is split, and the frequency of one of the two resultant beams is shifted by the ORFSL, which enables scanning from approximately DC upto the bandwidth. The phase noise of the CW laser is cancelled out when a microwave is generated. For mm wave ($\approx$ 30 GHz) \cite{Liu_2022}, two CW laser with a frequency offset are employed to add a frequency offset to the generated mm wave. Due to the use of two independent CW lasers, the phase noise of these lasers is transferred to the generated mm wave. For THz generation ($\approx$ 300 GHz) as demonstrated in this work, a significant frequency offset is added by using two optical tones from a DKS. Since the DKS operates in a mode-locked state, the phase noise of the THz wave (equivalent to relative phase noise between the comb modes) is far superior to the case where a THz generated from two CW lasers.

\begin{table*}
\centering
\caption{KEY PARAMETERS OF THE LINEAR FREQUENCY STEPPED SIGNALS}
\begin{tabular}{ c c c c c c}
 Ref  & Source of two tones & Center Frequency (GHz) & BW (GHz)    & Step frequency$\Delta f$ (GHz) & Max. N  \\ \hline
\cite{Lyu_2022}  & One CW laser & 10          & 20      & 0.04          & 500 \\   \hline
\cite{Schn_belin_2019} & One CW laser & 12.5       &25      & 0.085            &300    \\   \hline
\cite{Zhang_2022} & One CW laser & 15         & 30      & 0.1           & 200 \\   \hline
\cite{Zhang_2020multioctave} & One CW laser & 20   & 26      & 1, 3, 8, 10   & 29 \\ \hline
\cite{Liu_2022} & Two CW laser  & 34          & 20      &0.04/0.08      & 144 \\   \hline
This work      &  DKS  & 281           & 4.1    &0.083          & 50 \\   \hline

\end{tabular}
\end{table*}

\ifCLASSOPTIONcaptionsoff
  \newpage
\fi



%




\bibliography{scibib}

\begin{thebibliography}{10}
\newcommand{\enquote}[1]{``#1''}

\bibitem{Banks_2023}
P.~A. Banks, E.~M. Kleist, and M.~T. Ruggiero, \enquote{Investigating the
  function and design of molecular materials through terahertz vibrational
  spectroscopy,} {\protect\JournalTitle{Nature Reviews Chemistry}}  (2023).

\bibitem{Chen_2022}
X.~Chen, H.~Lindley-Hatcher, R.~I. Stantchev, J.~Wang, K.~Li, A.~H. Serrano,
  Z.~D. Taylor, E.~Castro-Camus, and E.~Pickwell-MacPherson, \enquote{Terahertz
  ({THz}) biophotonics technology: Instrumentation, techniques, and biomedical
  applications,} {\protect\JournalTitle{Chemical Physics Reviews}} \textbf{3},
  011311 (2022).

\bibitem{Nagatsuma_2016}
T.~Nagatsuma, G.~Ducournau, and C.~C. Renaud, \enquote{Advances in terahertz
  communications accelerated by photonics,} {\protect\JournalTitle{Nature
  Photonics}} \textbf{10}, 371--379 (2016).

\bibitem{Cooper_2011}
K.~B. Cooper, R.~J. Dengler, N.~Llombart, B.~Thomas, G.~Chattopadhyay, and
  P.~H. Siegel, \enquote{{THz} imaging radar for standoff personnel screening,}
  {\protect\JournalTitle{{IEEE} Transactions on Terahertz Science and
  Technology}} \textbf{1}, 169--182 (2011).

\bibitem{Lin2005}
I.~Lin, J.~McKinney, and A.~Weiner, \enquote{Photonic synthesis of broadband
  microwave arbitrary waveforms applicable to ultra-wideband communication,}
  {\protect\JournalTitle{{IEEE} Microwave and Wireless Components Letters}}
  \textbf{15}, 226--228 (2005).

\bibitem{Rashidinejad2014}
A.~Rashidinejad and A.~M. Weiner, \enquote{Photonic radio-frequency arbitrary
  waveform generation with maximal time-bandwidth product capability,}
  {\protect\JournalTitle{Journal of Lightwave Technology}} \textbf{32},
  3383--3393 (2014).

\bibitem{Danion_2014}
G.~Danion, C.~Hamel, L.~Frein, F.~Bondu, G.~Loas, and M.~Alouini, \enquote{Dual
  frequency laser with two continuously and widely tunable frequencies for
  optical referencing of {GHz} to {THz} beatnotes,}
  {\protect\JournalTitle{Optics Express}} \textbf{22}, 17673 (2014).

\bibitem{Kittlaus2021}
E.~A. Kittlaus, D.~Eliyahu, S.~Ganji, S.~Williams, A.~B. Matsko, K.~B. Cooper,
  and S.~Forouhar, \enquote{A low-noise photonic heterodyne synthesizer and its
  application to millimeter-wave radar,} {\protect\JournalTitle{Nature
  Communications}} \textbf{12} (2021).

\bibitem{hirata2005low}
A.~Hirata, H.~Togo, N.~Shimizu, H.~Takahashi, K.~Okamoto, and T.~Nagatsuma,
  \enquote{Low-phase noise photonic millimeter-wave generator using an awg
  integrated with a 3-d{B} combiner,} {\protect\JournalTitle{IEICE transactions
  on electronics}} \textbf{88}, 1458--1464 (2005).

\bibitem{Jerez_2019}
B.~Jerez, F.~Walla, A.~Betancur, P.~Mart{\'{\i}}n-Mateos, C.~de~Dios, and
  P.~Acedo, \enquote{Electro-optic {THz} dual-comb architecture for
  high-resolution, absolute spectroscopy,} {\protect\JournalTitle{Optics
  Letters}} \textbf{44}, 415 (2019).

\bibitem{parriaux2020electro}
A.~Parriaux, K.~Hammani, and G.~Millot, \enquote{Electro-optic frequency
  combs,} {\protect\JournalTitle{Advances in Optics and Photonics}}
  \textbf{12}, 223--287 (2020).

\bibitem{Jia_2022}
S.~Jia, M.-C. Lo, L.~Zhang, O.~Ozolins, A.~Udalcovs, D.~Kong, X.~Pang,
  R.~Guzman, X.~Yu, S.~Xiao, S.~Popov, J.~Chen, G.~Carpintero, T.~Morioka,
  H.~Hu, and L.~K. Oxenl{\o}we, \enquote{Integrated dual-laser photonic chip
  for high-purity carrier generation enabling ultrafast terahertz wireless
  communications,} {\protect\JournalTitle{Nature Communications}} \textbf{13}
  (2022).

\bibitem{Zhang_2019}
S.~Zhang, J.~M. Silver, X.~Shang, L.~D. Bino, N.~M. Ridler, and P.~Del'Haye,
  \enquote{Terahertz wave generation using a soliton microcomb,}
  {\protect\JournalTitle{Optics Express}} \textbf{27}, 35257 (2019).

\bibitem{tetsumoto2021optically}
T.~Tetsumoto, T.~Nagatsuma, M.~E. Fermann, G.~Navickaite, M.~Geiselmann, and
  A.~Rolland, \enquote{Optically referenced 300 {GH}z millimetre-wave
  oscillator,} {\protect\JournalTitle{Nature Photonics}} \textbf{15}, 516--522
  (2021).

\bibitem{Kuse_2022}
N.~Kuse, K.~Nishimoto, Y.~Tokizane, S.~Okada, G.~Navickaite, M.~Geiselmann,
  K.~Minoshima, and T.~Yasui, \enquote{Low phase noise {THz} generation from a
  fiber-referenced {K}err microresonator soliton comb,}
  {\protect\JournalTitle{Communications Physics}} \textbf{5} (2022).

\bibitem{ahn2005suppression}
T.-J. Ahn, J.~Y. Lee, and D.~Y. Kim, \enquote{Suppression of nonlinear
  frequency sweep in an optical frequency-domain reflectometer by use of
  {H}ilbert transformation,} {\protect\JournalTitle{Applied optics}}
  \textbf{44}, 7630--7634 (2005).

\bibitem{Herr_soliton}
T.~Herr, V.~Brasch, J.~D. Jost, C.~Y. Wang, N.~M. Kondratiev, M.~L. Gorodetsky,
  and T.~J. Kippenberg, \enquote{Temporal solitons in optical microresonators,}
  {\protect\JournalTitle{Nature Photonics}} \textbf{8}, 145--152 (2014).

\bibitem{kippenberg2018dissipative}
T.~J. Kippenberg, A.~L. Gaeta, M.~Lipson, and M.~L. Gorodetsky,
  \enquote{Dissipative {K}err solitons in optical microresonators,}
  {\protect\JournalTitle{Science}} \textbf{361}, eaan8083 (2018).

\bibitem{kovach2020emerging}
A.~Kovach, D.~Chen, J.~He, H.~Choi, A.~H. Dogan, M.~Ghasemkhani, H.~Taheri, and
  A.~M. Armani, \enquote{Emerging material systems for integrated optical
  {K}err frequency combs,} {\protect\JournalTitle{Advances in Optics and
  Photonics}} \textbf{12}, 135--222 (2020).

\bibitem{xiang2021laser}
C.~Xiang, J.~Liu, J.~Guo, L.~Chang, R.~N. Wang, W.~Weng, J.~Peters, W.~Xie,
  Z.~Zhang, J.~Riemensberger, J.~Selvidge, T.~J. Kippenberg, and J.~E. Bowers,
  \enquote{Laser soliton microcombs heterogeneously integrated on silicon,}
  {\protect\JournalTitle{Science}} \textbf{373}, 99--103 (2021).

\bibitem{kwon2022ultrastable}
D.~Kwon, D.~Jeong, I.~Jeon, H.~Lee, and J.~Kim, \enquote{Ultrastable microwave
  and soliton-pulse generation from fibre-photonic-stabilized microcombs,}
  {\protect\JournalTitle{Nature Communications}} \textbf{13}, 1--8 (2022).

\bibitem{Kuse_2020}
N.~Kuse, T.~Tetsumoto, G.~Navickaite, M.~Geiselmann, and M.~E. Fermann,
  \enquote{Continuous scanning of a dissipative {K}err-microresonator soliton
  comb for broadband, high-resolution spectroscopy,}
  {\protect\JournalTitle{Optics Letters}} \textbf{45}, 927 (2020).

\bibitem{kuse2021frequency}
N.~Kuse, G.~Navickaite, M.~Geiselmann, T.~Yasui, and K.~Minoshima,
  \enquote{Frequency-scanned microresonator soliton comb with tracking of the
  frequency of all comb modes,} {\protect\JournalTitle{Optics Letters}}
  \textbf{46}, 3400--3403 (2021).

\bibitem{Liu_2022}
Y.~Liu, Z.~Zhang, M.~Burla, and B.~J. Eggleton, \enquote{11-{GHz}-bandwidth
  photonic radar using {MHz} electronics,} {\protect\JournalTitle{Laser $\&$
  Photonics Reviews}} \textbf{16}, 2100549 (2022).

\bibitem{Chen_2018}
T.~Chen, W.~Kong, H.~Liu, and R.~Shu, \enquote{Frequency-stepped pulse train
  generation in an amplified frequency-shifted loop for oxygen a-band
  spectroscopy,} {\protect\JournalTitle{Optics Express}} \textbf{26}, 34753
  (2018).

\bibitem{Lyu_2022}
Y.~Lyu, Y.~Li, C.~Yu, L.~Yi, T.~Nagatsuma, and Z.~Zheng, \enquote{Photonic
  generation of highly-linear ultra-wideband stepped-frequency microwave
  signals with up to 6.10\textsuperscript{6} time-bandwidth product,}
  {\protect\JournalTitle{Journal of Lightwave Technology}} \textbf{40},
  1036--1042 (2022).

\bibitem{Zhang_2020multioctave}
Y.~Zhang, C.~Liu, K.~Shao, Z.~Li, and S.~Pan, \enquote{Multioctave and
  reconfigurable frequency-stepped radar waveform generation based on an
  optical frequency shifting loop,} {\protect\JournalTitle{Optics Letters}}
  \textbf{45}, 2038 (2020).

\bibitem{Schn_belin_2019}
C.~Schn{\'{e}}belin, J.~Aza{\~{n}}a, and H.~G. de~Chatellus,
  \enquote{Programmable broadband optical field spectral shaping with megahertz
  resolution using a simple frequency shifting loop,}
  {\protect\JournalTitle{Nature Communications}} \textbf{10} (2019).

\bibitem{Zhang_2022}
Z.~Zhang, Y.~Liu, and B.~J. Eggleton, \enquote{Photonic generation of 30 {GHz}
  bandwidth stepped-frequency signals for radar applications,}
  {\protect\JournalTitle{Journal of Lightwave Technology}} \textbf{40},
  4521--4527 (2022).

\bibitem{Ito_2005}
H.~Ito, T.~Furuta, F.~Nakajima, K.~Yoshino, and T.~Ishibashi, \enquote{Photonic
  generation of continuous {THz} wave using uni-traveling-carrier photodiode,}
  {\protect\JournalTitle{Journal of Lightwave Technology}} \textbf{23},
  4016--4021 (2005).

\bibitem{stone2018thermal}
J.~R. Stone, T.~C. Briles, T.~E. Drake, D.~T. Spencer, D.~R. Carlson, S.~A.
  Diddams, and S.~B. Papp, \enquote{Thermal and nonlinear dissipative-soliton
  dynamics in {K}err-microresonator frequency combs,}
  {\protect\JournalTitle{Physical review letters}} \textbf{121}, 063902 (2018).

\bibitem{kuse2019control}
N.~Kuse, T.~C. Briles, S.~B. Papp, and M.~E. Fermann, \enquote{Control of
  {K}err-microresonator optical frequency comb by a dual-parallel
  {M}ach-{Z}ehnder interferometer,} {\protect\JournalTitle{Optics Express}}
  \textbf{27}, 3873--3883 (2019).

\bibitem{shimizu1992technique}
K.~Shimizu, T.~Horiguchi, and Y.~Koyamada, \enquote{Technique for translating
  light-wave frequency by using an optical ring circuit containing a frequency
  shifter,} {\protect\JournalTitle{Optics letters}} \textbf{17}, 1307--1309
  (1992).

\bibitem{Kuse_2021}
N.~Kuse and K.~Minoshima, \enquote{Amplification and phase noise transfer of a
  {K}err microresonator soliton comb for low phase noise {THz} generation with
  a high signal-to-noise ratio,} {\protect\JournalTitle{Optics Express}}
  \textbf{30}, 318 (2021).

\bibitem{Wang_2022}
C.~Wang, Q.~Zhang, J.~Hu, C.~Li, S.~Shi, and G.~Fang, \enquote{An efficient
  algorithm based on {CSA} for {THz} stepped-frequency {SAR} imaging,}
  {\protect\JournalTitle{{IEEE} Geoscience and Remote Sensing Letters}}
  \textbf{19}, 1--5 (2022).

\end{thebibliography}

\bibliographystyle{osajnl}

%








\end{document}